\documentclass[preprint2,numberedappendix,tighten,twocolappendix]{aastex6}
\shorttitle{ECHO}
\shortauthors{Jacobs, et. al.}

\usepackage{amsmath}
\usepackage{graphicx}
\usepackage[figuresright]{rotating}
\usepackage{natbib}
\usepackage{ctable}
\usepackage{tabularx}
\usepackage{url}
\begin{document}

\title{The External Calibrator for Hydrogen Observatories}

\author{
Daniel C. Jacobs\altaffilmark{1,2},
Jacob Burba\altaffilmark{3},
Judd Bowman\altaffilmark{1},
Abraham R. Neben\altaffilmark{4}
Benjamin Stinnett\altaffilmark{1}
Lauren Turner\altaffilmark{1},
}

\altaffiltext{1}{School of Earth and Space Exploration, Arizona State University, Tempe, AZ, 85287}
\altaffiltext{2}{daniel.c.jacobs@asu.edu}
\altaffiltext{3}{Brown University, Providence, RI}
\altaffiltext{4}{MIT Kavli Institute, Massachusetts Institute of Technology, Cambridge, MA, 02139 USA}

\begin{abstract}
Multiple instruments are pursuing constraints on dark energy, observing reionization and opening a window on the dark ages through the detection and characterization of the 21~cm hydrogen line across the redshift spectrum, from nearby to z=25.  These instruments, including CHIME in the sub-meter and HERA in the meter bands, are wide-field arrays with multiple-degree beams, typically operating in transit mode.  Accurate knowledge of their primary beams is critical for separation of bright foregrounds from the desired cosmological signals, but difficult to achieve through astronomical observations alone.  Previous beam calibration work has focused on model verification and does not address the need of 21~cm experiments for routine beam mapping, to the horizon, of the as-built array.  We describe the design and methodology of a drone-mounted calibrator, the External Calibrator for Hydrogen Observatories (ECHO), that aims to address this need. We report on a first set of trials to calibrate low-frequency dipoles and compare ECHO measurements to an established beam-mapping system based on transmissions from the Orbcomm satellite constellation.  We create beam maps of two dipoles at a  9-degree resolution and find sample noise ranging from 1 to  2\%.  Assuming this sample noise represents the error in the measurement, the higher end of this range is roughly twice the desired requirement.  The overall performance of ECHO suggests that the desired precision and angular coverage is achievable in practice with modest improvements. We identify the main sources of systematic error and uncertainty in our measurements and describe the steps needed to overcome them.
\end{abstract}

\section{Introduction}\label{sec:intro}

A new generation of radio arrays is being developed that use large numbers of low-cost dipole antennas. This advance is made possible by developments in digital technology and enables exploration of new windows on the universe such as the epoch of reionization (EoR) via the redshifted 21 cm line \citep{Morales:2010p8093,Furlanetto:2006p2267,Madau:1997p2232}.  New telescopes in this regime include the MWA \citep{Tingay:2013p9022,Bowman:2013p9950}, PAPER \citep{Pober:2012p8800,2015ApJ...809...61A,2013ApJ...776..108J}, LOFAR \cite{Yatawatta:2013p9699}, HERA \citep{2016:deBoerHERAarxiv} and SKA-low \citep{Mellema:2013p10035,Mort:2016SKAlowimagingarxiv}.   Precise calibration of the primary beams of these dipole arrays has been found to be crucial to analysis of their observations.

The chief challenge in observing highly redshifted 21\,cm emission is discriminating the mK level spectral line signal against the $\sim$100K foregrounds.  Instrumental simulations have found that bright sources far from the pointing center, though attenuated significantly by the beam, introduce foreground signals which are highly chromatic and difficult to discriminate from the cosmological 21~cm spectral line signature \citep{Thyagarajan:2013p10039,2015ApJ...804...14T,Mort:2016SKAlowimagingarxiv}. These effects have been confirmed in both MWA and PAPER observations \citep{2015:ThyagarajanConfirmationwidefield,Pober:2016ApJ...819....8P}. 

Precise removal of foreground sources far into the sidelobes is one of the chief challenges of the primary analysis pipelines  \citep{2016:JacobsPipelinepaper}, however the amount which can be removed is limited by the accuracy of the knowledge of the antenna response.  More detailed electromagnetic models have been shown to improve the overall accuracy of flux reconstruction \citep{Sutinjo:2015RaSc...50...52S}.   However, electromagnetic models often cannot account for as-built variations between elements within an array, which has been shown to be significant and largely unavoidable without large added expense \citep{2016:NebenBeamformingerrors}.  

Some of the largest successes to date of improved antenna modeling accuracy have been in improving polarization response. Polarized signals have inherent spectral variation due to faraday rotation imprinted by magnetized ionized plasmas in the interstellar medium as well as the ionosphere. Leakage of these spectral signatures into the unpolarized reionization signal can be minimized with a combination of beam modeling and careful element design \citep{Jelic:2010p8293,Moore:2013p9941,Asad:2015LofarPol}.

\begin{figure*}[hbt]
\centering
\includegraphics[width=0.59\textwidth]{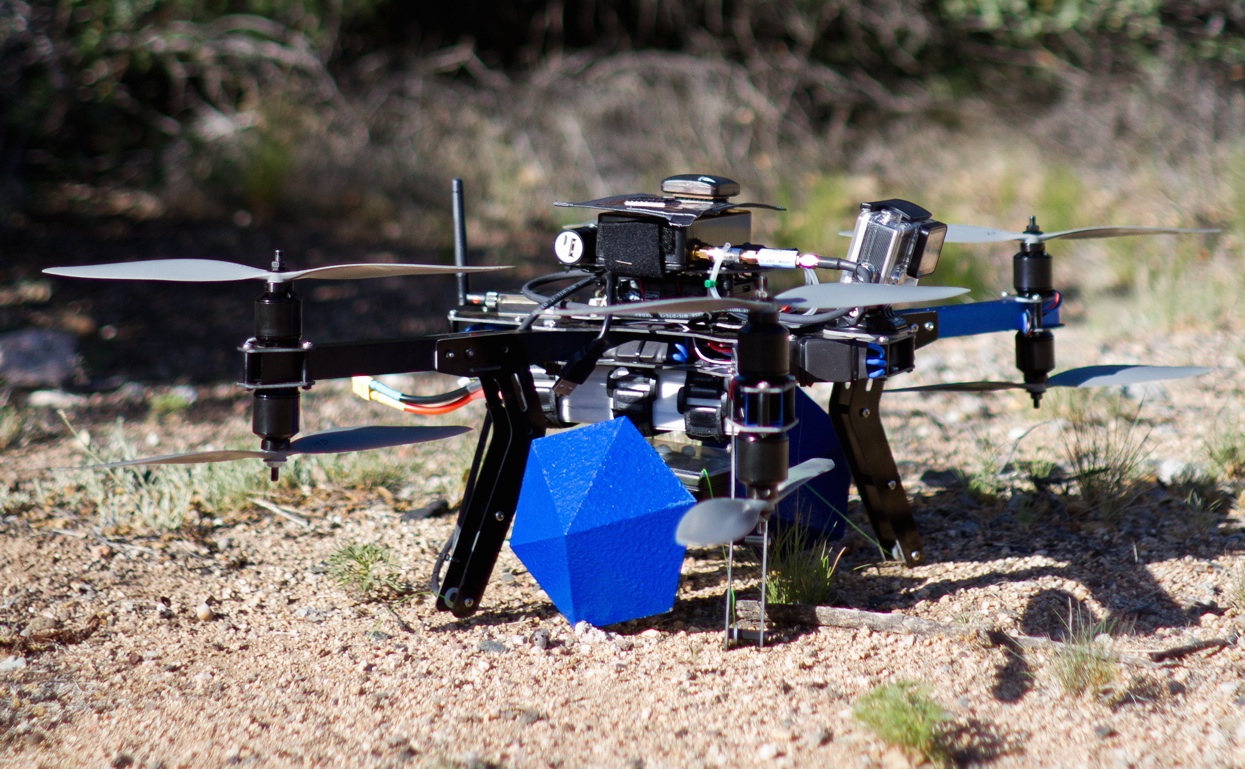}
\includegraphics[width=0.4\textwidth]{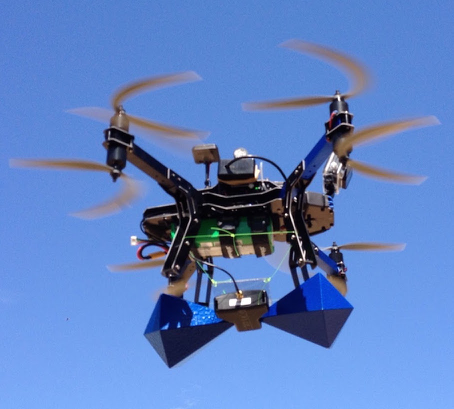}
\caption{3DR X8 octoquad drone used for ECHO prior to launch (left) and in flight (right).  The transmitter circuitry is contained in the black box mounted on top of the drone.  A copper plate below the box shields it from the electronics below and a GPS/magnetometer is attached atop the box. Beneath the drone, the blue BicoLOG transmitting antenna is slung with non-conductive monofilament. For scale, we note that the propellers are 30~cm tip-to-tip and the antenna cone has a diameter of 10~cm at its widest point.}
\label{fig:drone}
\end{figure*}

Another feature of 21~cm observations is the wide spectral bandwidth, for example, HERA, PAPER and MWA span fractional bandwidths of $\sim$2 from $\sim$100 to 200 MHz to cover the cosmological redshift range of interest. Successful foreground isolation necessitates knowledge of the beam response across this entire range.  Spectral variation of the beam has been found to cause otherwise smooth foregrounds to contribute more complicated contaminating structures \citep{2016:ThyagarajanBeamChromaticity,2016:EwallWiceHERAdisharXiv}. 

An inaccurate beam model also limits the accuracy of instrument calibration. The usual solution is to only use astronomical calibration sources in well-known parts of the beam, however incomplete foreground models have been shown to lead to contaminating spectral-line foregrounds \citep{2016:BarryCalibrationRequirements}.  Calibration error is not just limited to sky modeling.  Arrays such as PAPER and HERA are taking advantage of baseline redundancy to calibrate without sky models \citep{Liu:2010p10391,2014ZhengMITEoR,2015ApJ...809...61A}, however this technique is subject to error due to deviations from redundancy caused by beam variation from antenna to antenna. Though more work is necessary to understand the impact of beam non-redundancy, it is clear that more needs to be known about the actual as-built variations between antennas.  Accurate knowledge of the beam is also key during the design process as a validation of modeling and to explore design choices \citep{2014IAWPL..13..169V,2016:NebenHERAdish}. 

While instrument modeling and pipeline development results all agree that accurate beam characterization is critical, there is less certainty as to a specific requirement on that accuracy.  \citet{Shaw2015_chimemmodes} have placed a requirement of 0.1\% accuracy on the width of the beam for the CHIME experiment. This is roughly equivalent to the estimate within the HERA collaboration that 1dB standard deviation at the -40dB beam level is a good goal.

Taken together, we can summarize the 21\,cm needs for beam measurements as: 1) mapping of the in-situ / as-built antennas, 2) high measurement precision at the horizon, as well as the phase center, 3) maps at multiple frequencies across the operating band, 4) dense angular coverage in maps, and 5) full polarization sampling.

\subsection{Need for New Beam Calibration Techniques}

Beam calibration of low frequency dipole arrays poses several complications compared to traditional dish antennas. Most notable is that traditional beam calibration, where the beam is scanned across a bright isolated source, is not possible because arrays are fixed to the earth. Phased array beams are technically steerable but violate the assumption that the beam is constant as it scans. 

Drift scan calibration of dipole array beams is possible---observing point sources as they track across the beam---but it requires the test antenna/array to be embedded within an existing array that generates sufficient sensitivity to isolate a large number of radio sources to provide many tracks through the beam or availability of a large steerable dish with which to cross-correlate \citet{Berger:CHIME_beam_map2016-arxiv}.  For dipole arrays with large beams, drift scan calibration also requires stable instrument responses over hours, which may be difficult to achieve due to ambient temperature variations.  Even if source tracks can be measured at high sensitivity, there is still the issue that sky sources only measure right-ascension tracks along constant declination and cannot as easily constrain the beam in the orthogonal directional. \citet{Pober:2013p9942} have employed symmetry arguments to tie  different right-ascension tracks together, but this is not possible in general for more complicated dipole array beam patterns and is ultimately limited by the symmetry argument and in the sensitivity far out in the beam.   

Attempts have been made to use anechoic chambers to calibrate low-frequency phased arrays, but the measurements suffer since even the largest chambers cannot extend into the farfield pattern and the RF absorber material used in the chambers performs poorly below 150 MHz, creating reflections and resonances and, in any case, does not capture the as-built variation. Mapping beam responses of antennas in the field using the Orbcomm constellation of satellites has proven the most effective \citep{2015RaSc...50..614N,2016:NebenHERAdish} but is limited to only a single frequency (138 MHz).  

\subsection{Use of Drones}

Drone-based radio calibration has recently been explored for application to widefield 21~cm instruments. A drone-mounted calibration source was used to verify the accuracy of antenna response modeling for SKA-low stations \cite{2014IAWPL..13..169V} and then a second generation setup was then used as a phase calibrator \citep{2015ExA....39..405P}.  Fully mapping the beam of an antenna has been demonstrated by \citet{2015PASP..127.1131C}.  In that experiment, a 5m dish was mapped at 1~GHz with a noise source broadcast by a gimbal-mounted horn, providing a first demonstration of the drone beam-mapping concept and identifing places for improvement of the methods.   Future improvement areas they identified included: optimizing the flight path to better sample the beam, obtaining better drone positioning accuracy, improving characterization of the transmitter beam, and better modeling of the antenna under test (AUT). This last point was partially motivated by a lack of alternate beam mapping data with which to compare the results. 

Changes from the \citet{2015PASP..127.1131C} system must be made to shift from 1~GHz to lower frequencies used for reionization experiments.  Below $\sim$1~GHz, horns and other highly directive elements become prohibitively large and heavy for flight applications, motivating the use of electically compact antennas, such as dipoles like those used by \citet{2014IAWPL..13..169V}.  Fixing the transmitter antenna to the drone without a gimbal further reduces structural complexity and weight, but increases the need to understand the transmitter beam pattern since the orientation of the transmitting antenna will vary relative to the AUT.

Drone and other airborne calibrators are particularly well-suited to instruments with a reachable far-field distance.  The far field distance is defined as the distance at which wavefronts are planar across the element in question, an approximate estimate for the far field distance is the square of the aperture size over the wavelength. Table \ref{tab:table}  lists several representative 21~cm arrays and their approximate far field distances. The far field for HERA is $\sim$100m, which is a reasonable operating height for a small drone.  While desirable, we note that the far field threshold is flexible, as techniques exist to transform near field and Fresnel zone measurements into the far field \citep{johnson1973determination}.

\begin{deluxetable}{cccc}
\tablecaption{Far field distances for m and subm telescopes \label{tab:table}}
\tablehead{
\colhead{Element} & \colhead{Aperture Diameter} & \colhead{Wavelength} & \colhead{Far field} \\
\colhead{} & \colhead{m} & \colhead{m} & \colhead{m}\\
}
\startdata
PAPER & 2 & 2 & 2\\
MWA  & 5 & 2 & 12.5\\
HERA & 14 & 2 & 98 \\
LOFAR & 41 & 2 & 840.5\\
CHIME & 20 & 0.1 &  4000 \\  
HYRAX & 12 & 0.1 & 1,440 \\ 
VLA & 25 & 0.2 & 2,970\\
\enddata
\end{deluxetable}

In this paper, we report our progress towards overcoming the challenges of drone-based calibration with the development of the External Calibrator for Hydrogen Observatories (ECHO), which specifically addresses the 21~cm requirements identified above.  The ECHO system is described in detail in Section \ref{sec:design}.  It uses a broadband dipole as a transmitting antenna mounted in a fixed configuration beneath a multi-rotor drone. The antenna is driven by a sine-wave source that provides high transmission power at a tunable frequency. Beam maps are captured by flying the transmitter over the antenna using the HEALpix pixelation scheme to establish sampling waypoints.  As a test of the method, we have mapped two reference dipoles utilized in the beam-pattern measurement system of \citet{2016:NebenHERAdish} that uses the Orbcomm low-Earth orbit satellite constellation as transmission sources.  Comparing our measurements with the Orbcomm-derived beam patterns of the reference dipoles and to simulations, we quantify the measurement stability and overall accuracy of the ECHO system.  We discuss the results of our analysis in Section \ref{sec:data}, and in Section \ref{sec:conclusion} we summarize the conclusions from these tests and identify places for further improvement.

\begin{figure*}[htb]
\begin{center}
\includegraphics[width=0.6\columnwidth]{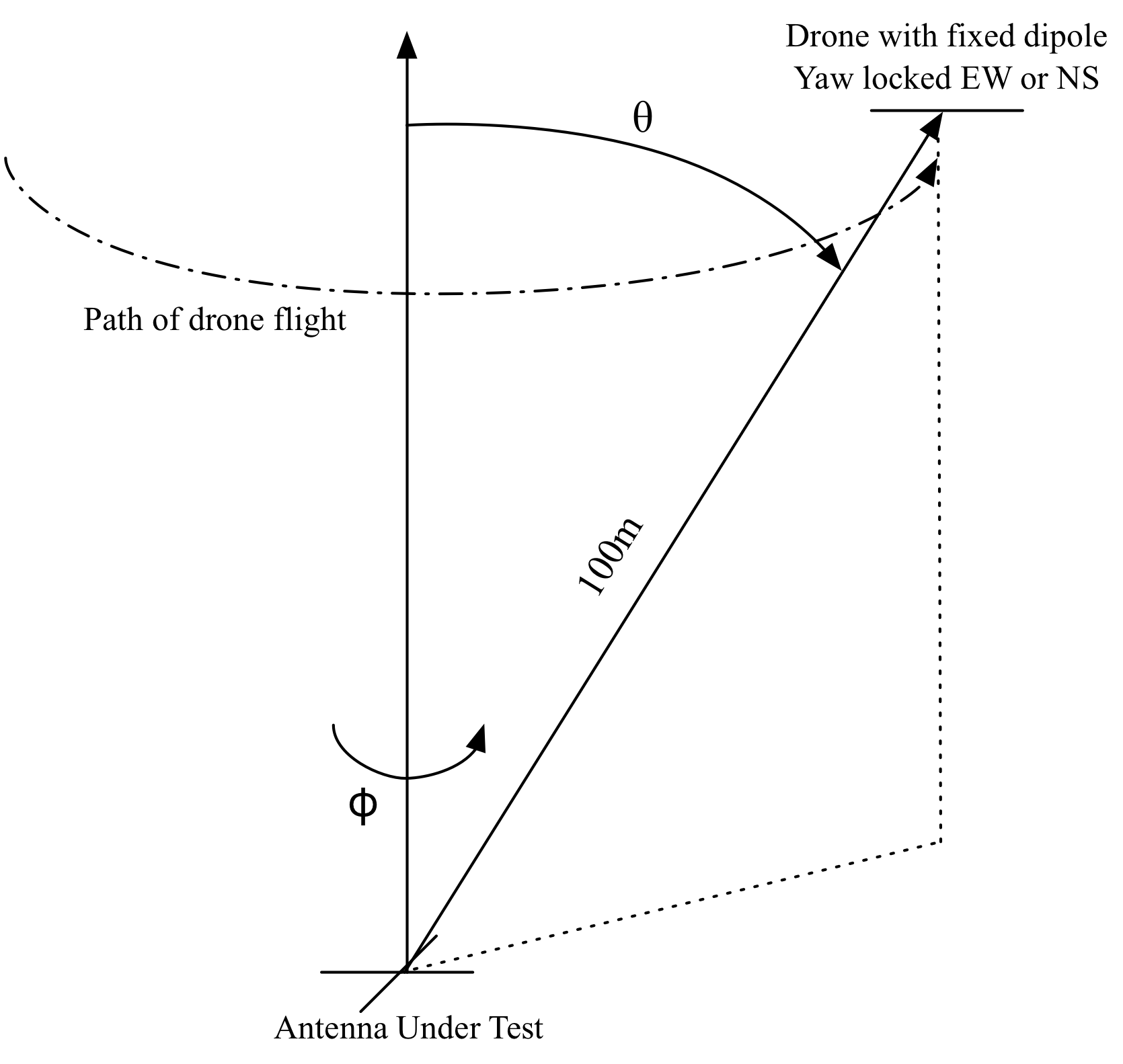}
\includegraphics[width=0.39\textwidth]{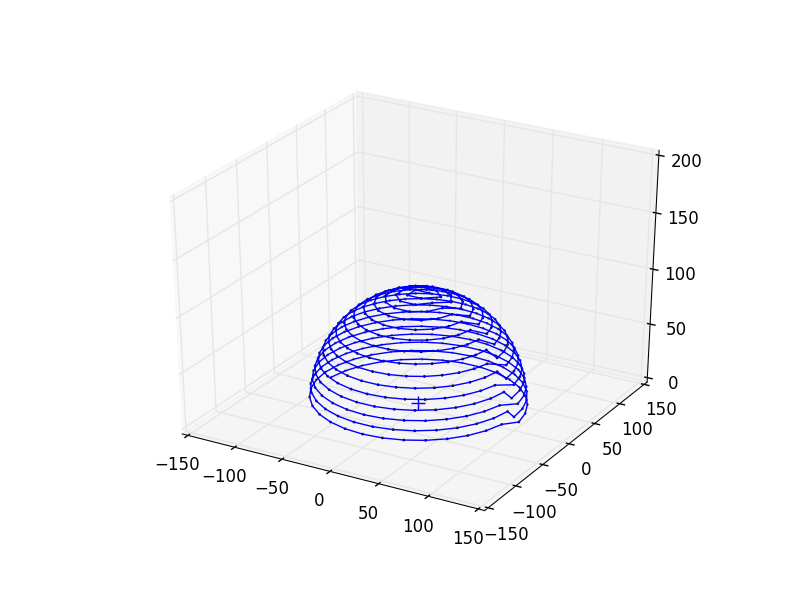}
\caption{Spherical coordinate system used in our anlaysis, defined relative to the AUT (left), and the programmed flight pattern which uses the HEALpix equal-area pixelization to form a spherical shell of waypoints with radius 100~m centered on the AUT (right). In all flights, the transmitting dipole antenna is mounted in a fixed position relative to the drone so that the drone yaw (rotation about the vertical) controls the transmitter polarization angle.  The drone is commanded to lock its orientation so that the transmitter is aligned either due north-south or east-west to match the orientation of the AUT dipoles. \label{fig:fligh_path}}
\end{center}
\end{figure*}

\section{Design and Method}
\label{sec:design}

\subsection{Drone and Transmitter}
We use an X8 octoquad from 3DRobotics (3DR) with a Pixhawk autopilot running Arducopter 3.5.  With a 10,000 mAh battery and a 1kg payload, this setup has a flight time of $\sim$15 minutes. Position is recorded by a Ublox GPS and a barometric altimeter, which together achieve a sub-meter position error.  The transmitter is a programmable Valon 5007 voltage controlled oscillator. Typically used as a clock signal, this device can produce a stable, continuous-wave signal tunable from 137\,MHz to 5\,GHz and is programmable via USB interface.  A continuous-wave signal is chosen over a broadband signal for our first testing because the power is concentrated in a narrow band and does not require additional amplification. This reduces both weight and complexity and enables better flexibility to avoid interference.  In the work reported here, the transmitter is programmed to broadcast at 137.5\,MHz with attenuation added to produce -22~dBm of output power.  This output power is chosen to reach a peak SNR of 40dB in our receiving station without evidence of saturation.  The transmit antenna is a BicoLOG 30100 biconical antenna manufactured by Aaronia. It is passive, has a smooth spectral response over an operational spectral range of 20\,MHz to 3\,GHz, and is light-weight yet robust.

\subsection{Transmitting Antenna Mount}

The antenna mounting is chosen to maximize the distance from the drone to facilitate modeling of the transmitting beam pattern. The X8 drone legs are too short to permit a fixed mounting further than two or three centimeters below the drone and the wingspan is too narrow for safe landing on longer legs. In our tests, the antenna has been suspended from the drone legs by non-conductive monofilament with a plastic bracket providing the filament/antenna mount point. On the ground, the antenna fits between the legs, in the air the antenna hangs 30\,cm below, as shown in Figure \ref{fig:drone}. The advantages of this system are that it is simple, extremely light-weight and provides isolation between drone and antenna. The disadvantages are that it lacks rigidity, stability, and because it must be re-tied after transport it is difficult to repeat precise alignment between tests, requiring additional beam modeling.  In this scheme, with the antenna fixed in position relative to the drone, the drone yaw (rotation about the vertical) controls the transmitter polarization angle. 

\subsection{Flight Path}

One of the main difficulties encountered by \citet{2015PASP..127.1131C} was in matching the flight pattern to the size scale of the variation in the beam.  They chose a cartesian grid pattern with a separation such that the drone passed through the primary lobe roughly three times.  When combined with the on/off switching they used to subtract the 1~GHz noise of the drone motors, the number density of samples became extremely limited and led to spurious results when interpolating to make a complete beam map.  Here, our goal is to smoothly map the beam of a very wide-field element from zenith to the horizon, which would be difficult with a constant altitude flight pattern.  This is also true of the analysis pipelines where the distortion caused by the gnomic projection to a tangent plane makes for a poor gridding scheme at the horizon. 

In many 21~cm pipelines, the HEALPix pixelization scheme \citep{Gorski:2005p7667} is used to store images and beam models.  HEALPix divides the sky into equal area pixels with the pixel count and resolution selectable by powers of two. Pixels can be ordered either by their position in the hierarchical doubling (NEST) or in a spiral order with  longitude $\phi$ fastest (RING).  We chose a flight pattern which follows the NEST pattern to give us a spherical shell flight pattern centered on the receiving antenna with a chosen radius. Following a conservative safety practice limiting our drone flights to within a good line of sight of the operator, and considering the FAA flight ceiling of 400 feet, we set the radius (and thus the maximum altitude) to 100~m~(328~feet). The NSIDE parameter sets the number of waypoints and, effectively, the resolution of the beam sampling.  This number is constrained, ultimately, by the amount of time available to create a beam map---at constant speed, a denser grid will take longer to record.  The slowest variable to be sampled in our telemetry is GPS position at 2 Hz. We've chosen to fly at 1~m/s, hence the drone will acquire about three samples per angular degree with a HEALpix NSIDE of 8.  At this resolution and speed, the full $2\pi$ steradians of a beam response above the horizon can be mapped to a resolution of 9~degrees in 60~minutes with four 15-minute sorties.  Figure~\ref{fig:fligh_path} illustrates the ECHO coordinate system and flight path.

The polarization angle of the antenna is kept fixed by inserting a Region of Interest (ROI) command into the beginning of the waypoint program.  The ROI command causes the drone to maintain a fixed heading in the direction of a set location which we choose to be a distant location due East or North to match the polarization of the antenna being mapped.

\subsection{Antennas Under Test}

We tested ECHO by mapping two antennas that are used in the Orbcomm-based beam mapping system developed by \citet{2015RaSc...50..614N,2016:NebenHERAdish} at Green Bank Observatory\footnote{The Green Bank Observatory \emph{was} operated by the National Radio Astronomy Observatory which is a facility of the National Science Foundation operated under cooperative agreement by Associated Universities, Inc.}.  The Orbcomm system measures the power received from multiple satellites, operating in narrow band channels near 137 MHz, taking advantage of the telemetry in Orbcomm signals to associate the received channel powers with the orbital positions of the satellites.   The time variation of the digital signal is removed by simultaneously measuring the power received by an AUT and a nearby, dual polarization, sleeved reference dipole over a 2m x 2m wire mesh ground plane.  In the Orcbomm analysis, the ratio between the received power from the AUT and reference dipole is insensitive to satellite transmitter variations. The reference dipole acts as a precision calibration element and can be calibrated out using a simple beam model.  The accuracy and repeatability of this scheme can be established with a ``null test'' where the AUT is replaced with a second identical reference dipole.  

\begin{figure}
\begin{center}
\includegraphics[width=\columnwidth]{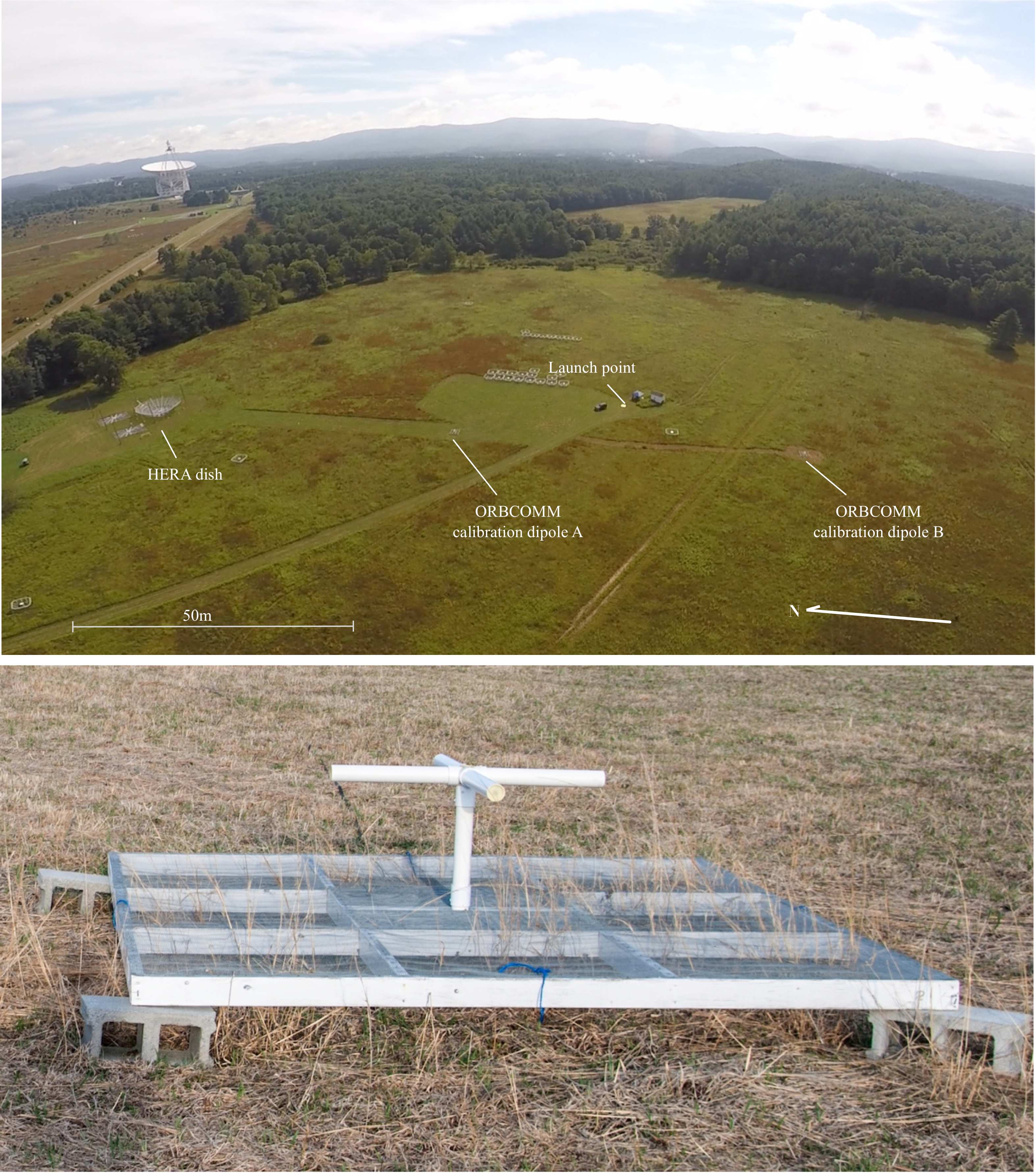}
\caption{Aerial view of the Orbcomm experimental setup in Green Bank (top) used for the ECHO test.  The two AUTs used for ECHO are reference antennas for the Orbcomm system, separated by 100~m on a north-south line. Each antenna (bottom) is a dual polarization sleeved dipole mounted on a 2m~x~2m mesh set $\sim$20~cm above the ground.  Two spherical flight patterns were performed by ECHO on each antenna to record received power for the transmitter aligned NS and EW.  }
\label{fig:GB_aerial}
\end{center}
\end{figure}

The Orbcomm system was in this ``null test'' configuration, shown in Figure \ref{fig:GB_aerial}, in preparation for mapping the HERA beam \citep{2016:NebenHERAdish} when we performed the ECHO tests described here.   Hence, we used the system's two reference dipoles as our AUTs for ECHO.  In the rest of this paper, we will refer to the two AUTs as antenna A and B, respectively, and the two dipole polarizations of each antenna as east-west (EW) and north-south (NS).

\subsection{Receiver}

Received power by our AUTs was collected using the spectrometer setup of the Orbcomm satellite mapping array as described in \citet{2016:NebenHERAdish}, their Section 2.4. The software was modified  to increase the spectral dump rate and allow real-time inspection of the data, but otherwise the system was operated just as it was for Orbcomm-based beam mapping, with spectra recorded from our two AUTs at a time cadence of 4~Hz and a spectral resolution of 2~kHz.   Although spectra were recorded simultatenously from both AUTs, we made separate flights centered on each AUT individually for the measurements reported here.  We do not analyze data from an AUT not at the center of a flight pattern.

\begin{figure}[htb]
\includegraphics[width=\columnwidth]{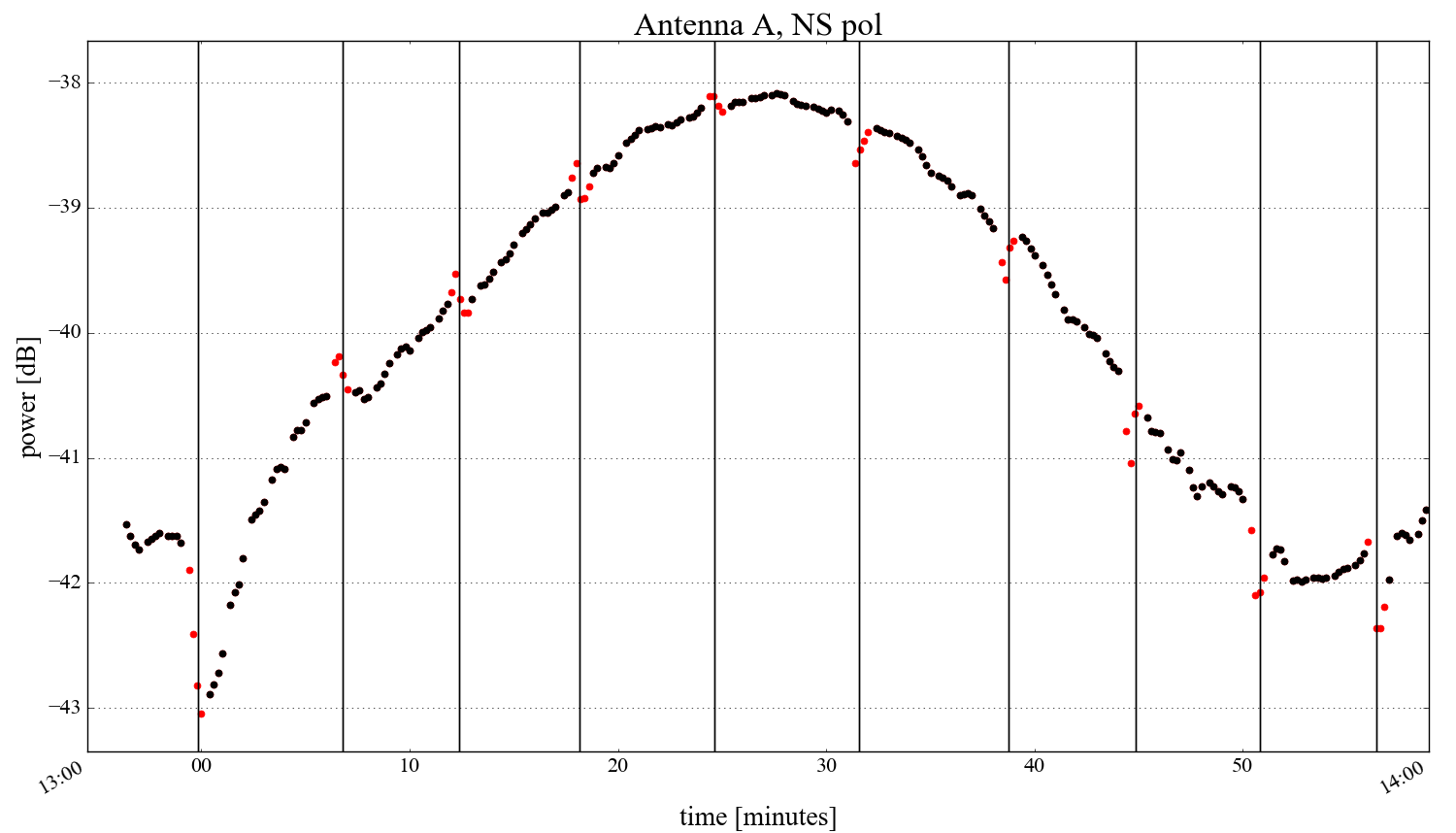}
\caption{Received power as a function of time. The drone is observed to shudder slightly when it goes through waypoints (vertical lines) an effect visible in the received power. We flag received data within 0.5~seconds of a waypoint arrival (points in red).}
\label{fig:waypoint_flagging}
\end{figure}

\subsection{Data processing}
\label{sec:processing}

We generate beam maps by interpolating the received power trace at times when recorded GPS positions are available.  Careful observation of the autopilot behavior and inspection of logs suggests that the X8/Arducopter combination suffers from occasional stability `hiccups'. Finding that these are characterized primarily by outliers in the yaw, we have flagged points greater than 2~$\sigma$ above the mean yaw.  Another slight yaw outlier is observed to occur as the drone passes through each waypoint.  These are often accompanied by visible bumps or dips in the received power trace, as can be seen in Figure~\ref{fig:waypoint_flagging} near each of the vertical lines marking a waypoint passage.  Points within half a second of each waypoint have been flagged.

\begin{figure*}[ht]
\begin{center}
\includegraphics[width=\textwidth]{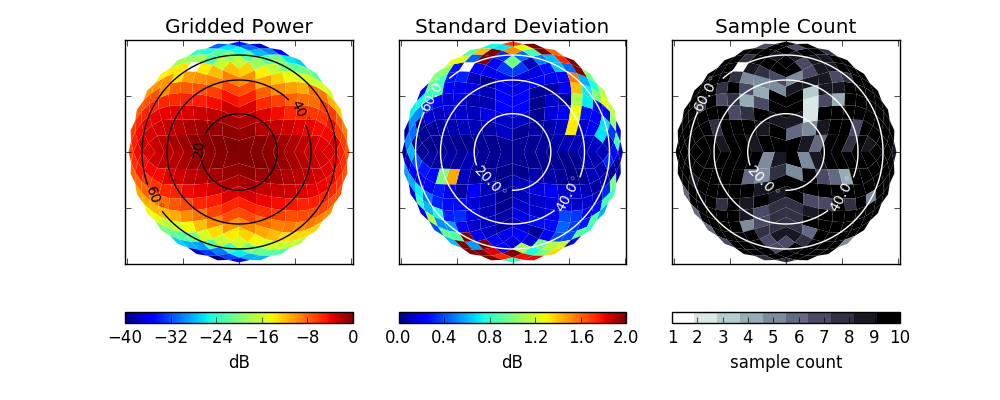}
\caption{(left) Received power pattern for the NS dipole of antenna A.  The pattern is shown in its raw form before it has been normalized to remove the effect of the transmitting antenna beam.  The pattern is effectively the product of two simple dipole patterns and exhibits the expected broad-beam response with nulls at the north and south (top and bottom) edges.  It is plotted on the same HEALpix grid as used for the flight waypoints.   (middle) The standard deviation of samples falling within a given pixel of the received power map. The standard deviation is calculated relative to the average in dB units, not linear units.  (right) The number of samples within in a given pixel of the map. Most pixels contain four or more samples.}
\label{fig:beam_std_count}
\end{center}
\end{figure*}

After flagging, data are gridded using a nearest pixel approach into a HEALpix grid using an NSIDE of 8, the same pixel scheme as used to create the flight pattern.  The gridder records the mean, standard deviation, and total sample count of all the samples falling into each pixel. An example gridded pattern is shown in Figure~\ref{fig:beam_std_count}.

\begin{figure}
\includegraphics[width=\columnwidth]{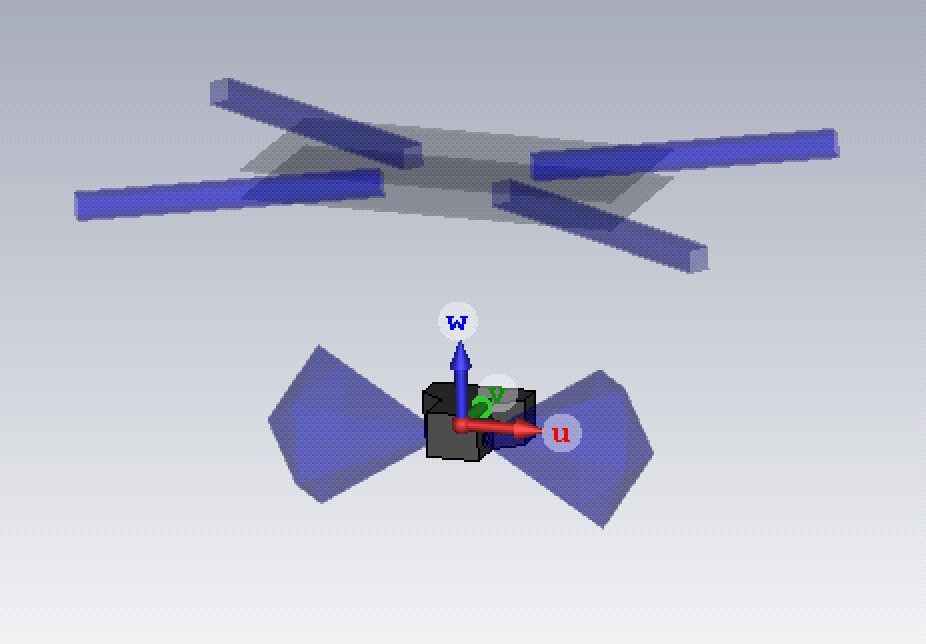}
\caption{CST model of the transmitting antenna suspended beneath the metal arms of the drone. The central body, shown in transparent black, is made of non-conductive materials. The largest metallic objects not modeled are the battery which is mounted beneath the central core, the motors on the ends of the arms, and the electronics mounted above the central body.}\label{fig:tx_cst}
\end{figure}

The measured power pattern is a product of the power patterns of the transmitter beam and receiving AUT beam. To remove the effect of the transmitter beam, we divide out by a simulation of the transmitter beam created in CST Microwave Studio based on a mechanical model of the drone plus antenna. The model is shown in Figure~\ref{fig:tx_cst} and includes the biconical transmitter antenna, as well as the metal arms of the drone.  The central body of the drone (shown as a translucent grey rectangle) is made of plastic and carbon fiber and contains control electronics, batteries, and the transmitter circuitry. These components are not modeled, nor were the motors on the ends of the arms.   

\section{Data}
\label{sec:data}

Data were collected in a series of four mapping runs in August, 2015 using the Orbcomm setup at Green Bank Observatory.   Historical spectral monitoring from the site was examined to identify a channel close to the Orbcomm transmission channels, but rarely used, yielding the transmission frequency selected for ECHO of 137.5~MHz.  The spectrum was monitored throughout operations for obvious signs of an overlapping Orbcomm transmission.  Each our two AUTs was mapped twice, once with the ECHO transmitter polarization fixed in the east-west direction and once north-south.  The resulting four data sets were flagged and gridded as described in Section \ref{sec:processing}.  A representative map is shown in Figure~\ref{fig:beam_std_count}.

\subsection{Agreement with Model}

\begin{figure}
\includegraphics[width=\columnwidth]{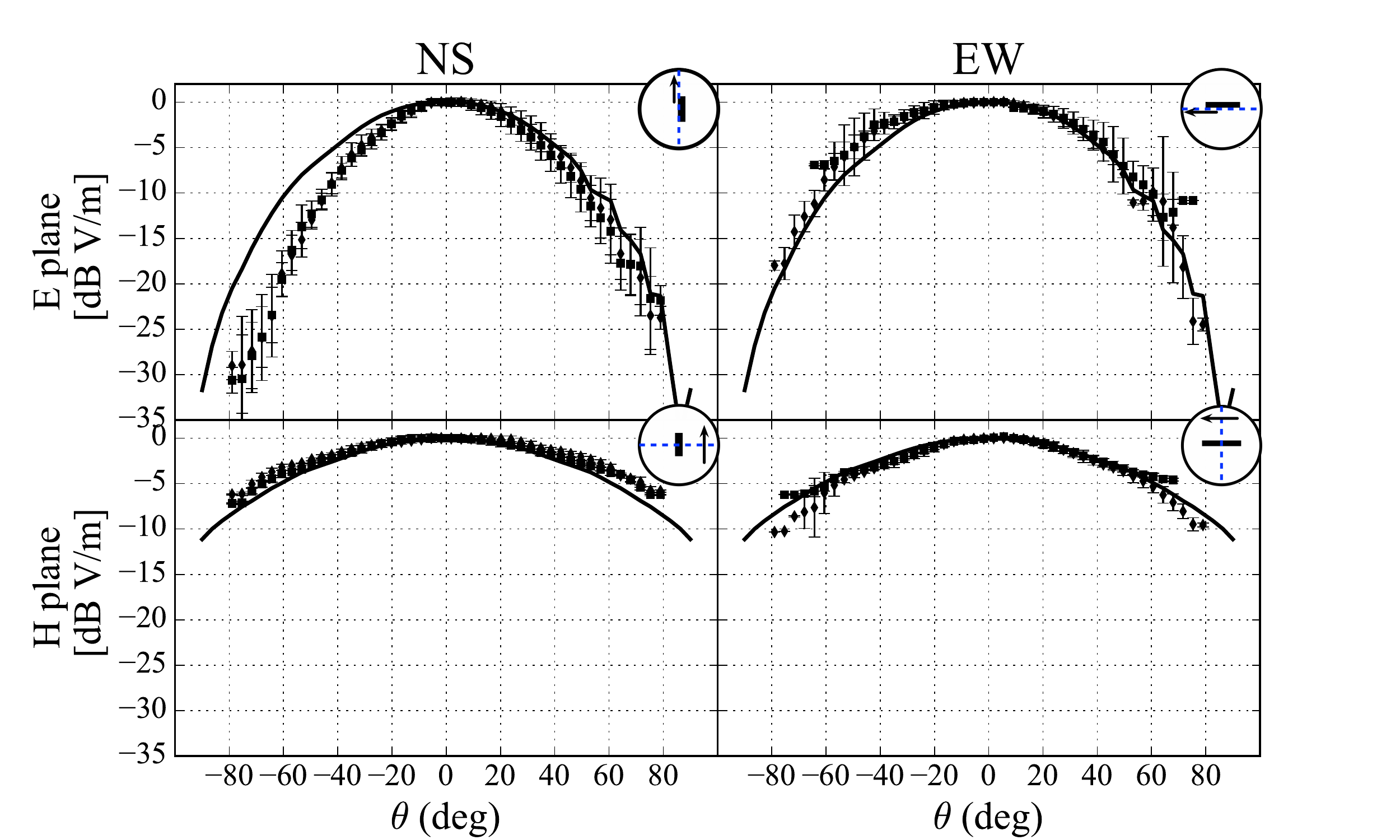}
\caption{Cross-sections of AUT beam maps after normalizing by the transmitter beam pattern.  Antenna A is square symbols, antenna B is diamonds, and the simulated beam model is the solid black line.  ECHO measurements of the antenna A and B beam patterns agree well with each other to within the empirical sample noise errorbars---even out to the lowest elevations sampled at $|\theta|\sim80$\arcdeg{} and weakest resopnses at 30~dB below peak.  The antennas also agree well with the model for the EW dipole polarization, but differ significantly from the model in the NS case.  We expect the EW and NS polarizations to have the same beam patterns relative their respective E- and H-planes, hence the ECHO measurements suggest that the AUT beam model might not fully capture the antenna as built.   \label{fig:GB_slices_quad}}
\end{figure}

In Figure \ref{fig:beam_std_count}, we can see the basic features of a dipole with its polarization axis oriented north-south above a ground plane: a nearly uniform response along the east-west direction and deep nulls to the north and south.  In the nulls beyond $\sim$70\arcdeg{} from the zenith, the amplitude falls more than 30~dB below the peak response.  The standard deviation between measurements within a gridded pixel is largest near the nulls, reaching 2~dB, compared to $\lesssim$0.4~dB in the main part of the beam.  We take this sampling noise as a reasonable estimate of the system's uncertainty in the discussion below.  

We compare the measured beam patterns against an electromagnetic model in order to examine each of the four measurements independently and assess the accuracy of the measurements.  As sleeved dipoles above small ground planes, antennas A and B are relatively straightforward to simulate in modern codes such as CST, FEKO, and HFSS. Here we use the CST model constructed by \citet{2015RaSc...50..614N} for use in the Orbcomm experiment.  The model includes one active polarization with the orthogonal polarization as a passive element, the ground screen, and a meter margin of surrounding earth. 90\arcdeg{} symmetry is assumed for the second polarization.   In Figure~\ref{fig:GB_slices_quad}, we plot E and H-plane cross-sections through the beam patterns measured by ECHO, as well as cross-sections for the simulated beams for the AUTs.   The cross-sections cut through the maps along the E-plane (parallel to the receiving dipole) in the top panels, and the H-plane (perpendicular to the receiving dipole) in the bottom panels.  The E-plane passes through the the dipole nulls, while the H-plane passes through the part of the beam with the flattest response.   The right column shows the cross-sections for the EW polarization of the AUTs.  In this column, we see that the beams of the two AUTs agree to within the empirical sample noise errorbars in most cases.  Agreement with the model is also within the sample noise errorbars.  The antenna B measurements were peformed to lower elevations than those for antenna A and we see that the agreement with the model is good even for the lowest elevation points with antenna B, which reached $\theta=\pm80$\arcdeg{} and a received power 25~dB below the peak.    

The left column of Figure \ref{fig:GB_slices_quad} shows cross-sections from the beam measurements of the NS polarization of the AUTs.  We note that the measured beam pattern for the NS polarization of antenna A has been corrected for a misalignment of the transmitting antenna, which we will discuss in detail below.  Measurements of antenna B were unaffected.   Here, we can see that antennas A and B agree well with each other, even at received power levels of 30~dB below peak response.  However, neither antenna agrees well with the model in the NS polarization.  There is a systematic offset between both AUTs and the electromagnetic model, particularly in the E-plane for $\theta<0$\arcdeg{}.  The NS polarization cases have deeper nulls in the E-plane and a flatter response along the H-plane than the EW cross-sections.   Due to the symmetry of the AUTs, we would expect the EW and NS polarizations to yield identical beam cross-sections in their respective E and H-planes since we always align our transmitter antenna to the active dipole axis.  Hence, it is surprising to see any discrepancy between polarizations.  Since both antennas exhibit the same behavior, it suggests the issue is with the antennas themselves, and is not stability issue with the ECHO system or a systematic error in the modeled transmitter beam pattern.  We investigated if a simple rotation of the NS arms relative to the EW arms within the AUTs can account for this difference and found that it cannot.  Another possibility is that an unmodeled difference in the construction of the NS dipoles compared to the EW may cause the beam model to be innaccurate.  Small geometrical assymetries near the connection between a dipole and its active balun are known to cause beam distortions and we speculate that something of this nature may be the case in the NS polarization dipoles.

\subsection{Identification of Misaligned Transmitter Antenna}
\label{sec:orbcomm}

\begin{figure}
\includegraphics[width=\columnwidth]{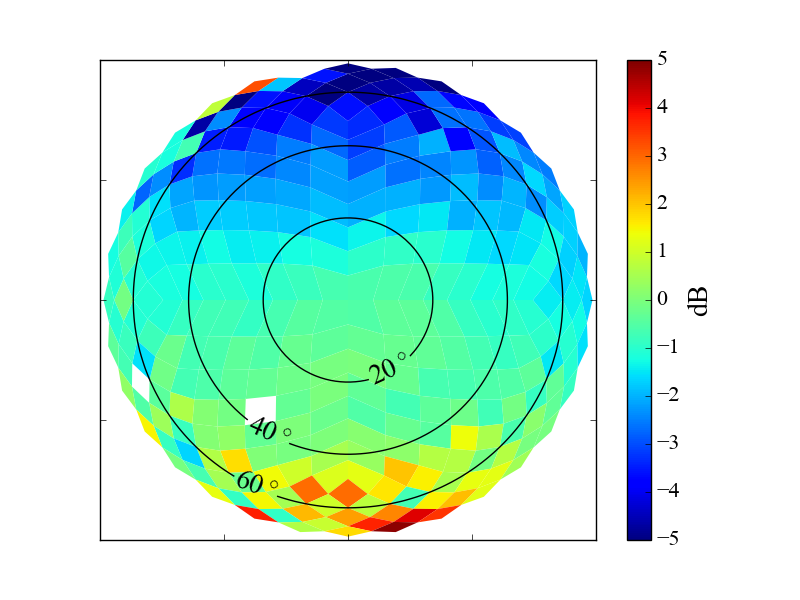}
\caption{Ratio between the antenna A and B beam patterns for the NS polarization.  The ratio is expected to be close to unity (0~dB), with deviations revealing measurement errors or inconsistencies in the two AUTs. Here, we see the presence of a strong north-south gradient, which we have identified with a 10\arcdeg{} misalignment (tilt) of the transmitting antenna during the mapping run of the NS polarization of antenna A.  The misalignment has been corrected for in Figures~\ref{fig:GB_slices_quad},~\ref{fig:GB_ratio_maps}, and~\ref{fig:GB_ratio_slices} by normalizing the affected raw received power pattern with a recalculated transmission beam pattern. }\label{fig:GB_NS_ratio_uncalibrated}
\end{figure}

\begin{figure*}
\begin{minipage}{0.45\textwidth}
\centering
\includegraphics[width=\textwidth]{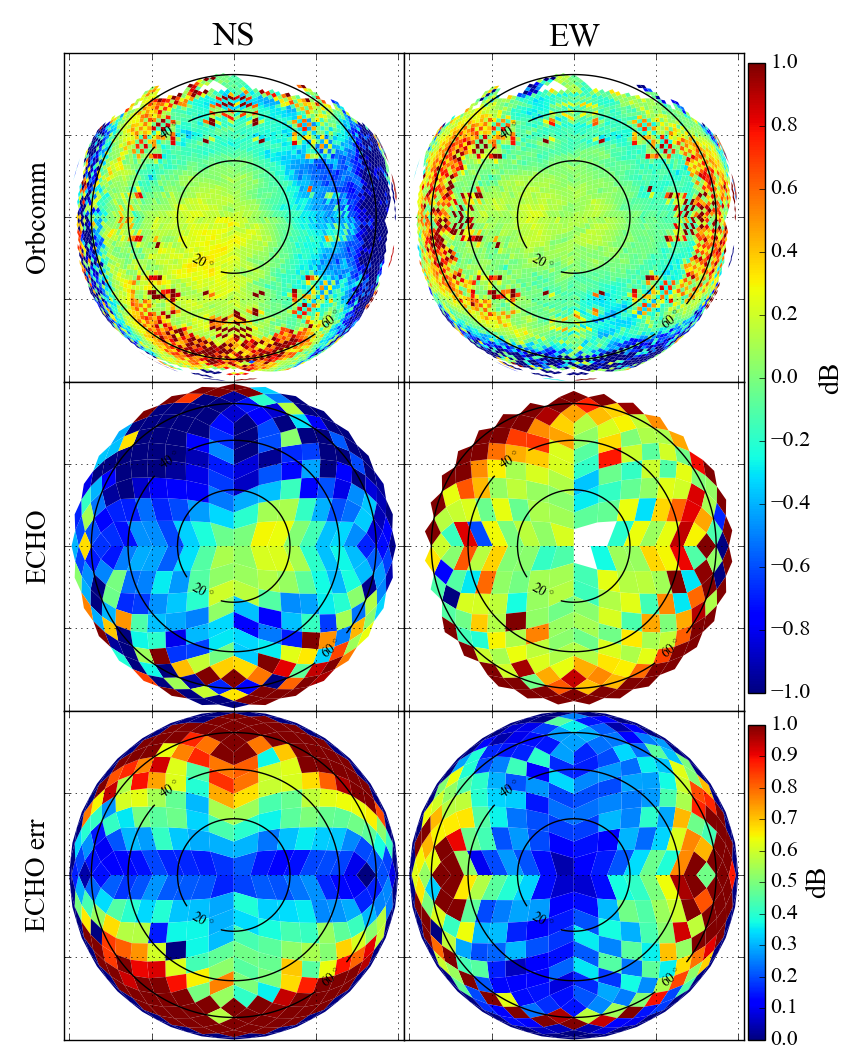}
\caption{Beam pattern ratios between antennas A and B (middle panels) compared against similar measurements made with the Orbcomm satellite system (top panels).  Both methods agree that the EW dipoles are more alike than the NS, though disagree on the nature and extent of the difference.  The bottom panels show the ECHO sample noise within each pixel found by propagating the standard deviation of the individual maps used in the ratios. The error is smallest in the H-plane and increases towards the nulls, this is consistent with degree-scale rotation instability in the drone-mounted antenna.}\label{fig:GB_ratio_maps}
\end{minipage}
\centering
\begin{minipage}{0.45\textwidth}
\centering
\includegraphics[width=\columnwidth]{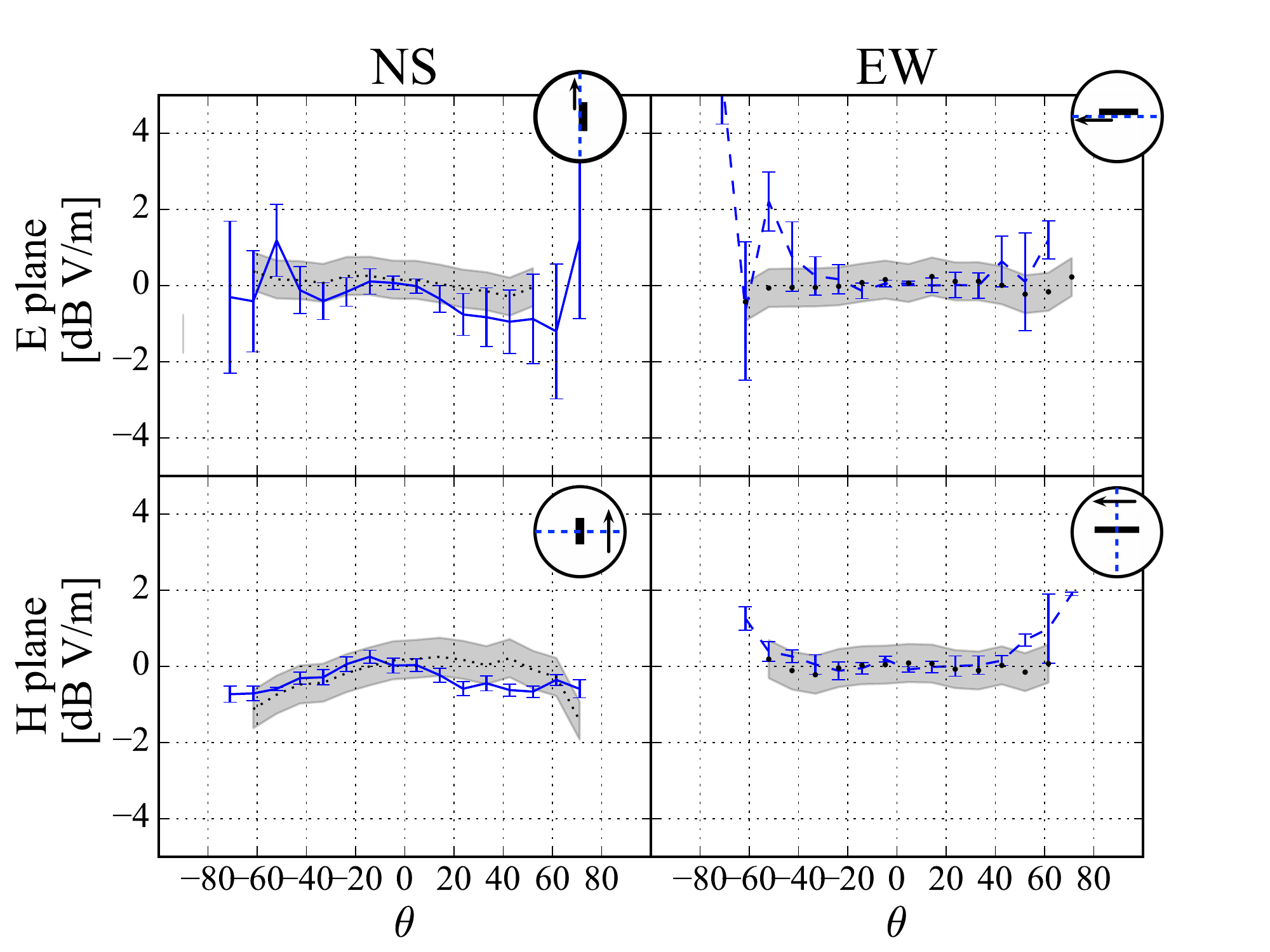}
\caption{Cross-sections of the beam patterns plotted in Figure~\ref{fig:GB_ratio_maps}.  ECHO ratios are shown in blue and Orbcomm in gray.     Assuming perfectly identical receiver antennas, this ratio would be unity (0~dB). Here, we see deviations generally on the scale of the emperical sample noise errorbars.  Errors are larger in the E-plane where the transmitting and receiving dipoles are aligned on their nulls and small rotations in the transmitter orientation that are not accuractely accounted for can cause larger effects in the received power.  Orbcomm and ECHO agree well in most cases.
\label{fig:GB_ratio_slices}}
\end{minipage}\hfill
\end{figure*}

We took the ratio of the antenna A beams to the antenna B beams, treating the NS and EW dipole polarizations separately.  Assuming the ECHO system is stable between mapping runs, the ratios should be unity.  In practice, any instability in ECHO or differences between the two antennas due to construction irregularities or nearby environmental properties will result in ratios that differ from unity.  As a reference, we derived the same ratios between the antenna beams from data collected nearly simultaneously during an Orbcomm ``null test''.   

Comparing the ECHO and Orbcomm-derived beam ratios, we found that the beam patterns measured by ECHO for the EW dipoles yielded ratios similar to those derived from the Orbcomm measurements, but the raw beam patterns measured by ECHO for the NS dipoles displayed a systematic slope in the NS direction, as evident in Figure~\ref{fig:GB_NS_ratio_uncalibrated}.  A systematic slope in the measured beam ratio of the two AUTs that is aligned with the polarization of the transmitter can be produced by a modest tilt in the transmitting antenna or one of the receiving antennas.  Since the Orbcomm data do not show a similar feature, we assume the slope is due to a tilt in the ECHO transmitting antenna.  Such a tilt is a likely failure mode of the monofilament transmitter suspension method.  

If stable within any given mapping run, the tilt can be accounted for during our data reduction by modifying the transmitter beam model used to normalize the data.  Inspection of photos taken during the field trials revealed several instances of small antenna mounting tilts, mostly in the last mapping run of the experiment, on the NS arm of station A.   We find that a rotation of the transmitter antenna by $\Delta \theta = 10$\arcdeg{} fits the observed slope well while staying within the bounds of likely misalignments observed in the field and in the photos.  To compensate for the slope, we have applied a revised transmitter beam model that accounts for the misalignment of the transmitting antenna during the ECHO measurements of the NS dipole of antenna A.  An unrotated transmission beam is used to normalize  the other beam measurements.  This correction has been applied to the data shown in Figures~\ref{fig:GB_slices_quad},~\ref{fig:GB_ratio_maps}, and~\ref{fig:GB_ratio_slices}.

The ratios between the measured antenna beam patterns are plotted in Figures~\ref{fig:GB_ratio_maps} and~\ref{fig:GB_ratio_slices}.   Comparing Figure \ref{fig:GB_NS_ratio_uncalibrated} to the beam-tilt corrected version plotted in the middle panel of Figure~\ref{fig:GB_ratio_maps}, we can see that the original 10~dB slope across the beam has been reduced to a much smaller difference of  $\sim$0.5~dB, which is similar to the typical sample noise in the ratio maps, as shown in the bottom panel of Figure~\ref{fig:GB_ratio_maps}.   

The ratio maps show again that ECHO measurements of antennas A and B exhibit nearly identical beam patterns to within the sample noise, except at large zenith angles.  This discrepancy between the beam patterns of antennas A and B at large $|\theta|$ is especially evident in the EW polarization. At large zenith angles, the ECHO drone is close to the ground.  It is likely that ground propagation effects influence the transmitted signal at large $|\theta|$, but are not accounted for in our analysis presently.    Overall, the Orbcomm and ECHO data agree that the NS polarizations are not as similar to each other as the EW polarizations for the two antennas.

\subsection{Accuracy and Source of Sample Noise}

Figure~\ref{fig:GB_ratio_slices} shows that the typical sample noise in the ratio maps is less than 1~dB, but climbs to 3~dB at the beam nulls.  Any deviation between the ratios derived from the ECHO and Orbcomm measurements is typically within this level of measurment variation.  Taken in conjunction with the agreement between the ECHO beam patterns and the simulated AUT beam patterns, the agreement between the ECHO ratio maps and the Orbcomm ratio maps suggests that the ECHO system yields accurate measurements of the AUTs.  In percentage terms, the lowest sample noise is in the H-plane, where the deviations between samples within a given pixel were typically 1\%. These variations are roughly on the same scale as the measured difference between the two antennas, as well as the difference between modeled beam patterns and data.  

Sample noise increases approximately linearly with azimuthal angle and reaches a maximum of 6\% in the E-plane at high zenith angles, where the AUT and transmitter dipoles have their nulls aligned.  Around the nulls, small rotations of the transmitter antenna can cause large changes in the received power.  If these rotations are not accounted for in the drone orientation telemetry, they will yield errors in the beam patterns.  Applying a gaussian random rotation to the transmitter beam model, we find that the observed fractional variation near the nulls is consistent with a transmitter polarization  rotation RMS of 1.3\arcdeg{}.  This is not far from the observed RMS of the drone yaw of 2.2\arcdeg{}.  The measured yaw is likely to be influenced by other errors such as magnetic interference from the motor currents and is likely an overestimate of actual drone motion.

\section{Discussion}
\label{sec:conclusion}

In this paper, we've described the development and characterization of an airborne radio calibrator specifically targeted at the needs of high-redshift 21~cm experiments. The necessity for precision foreground subtraction at one part in 10,000 motivates our desire to make a high fidelity map of the polarized beam response of the as-built antennas. Simulations have shown that beam maps must cover the full $2\pi$ steradians of the sky at uniform precision. 

The design studied here is a relatively simplistic execution with a single dipole fixed beneath a small, low-cost, commercially-available, multi-rotor drone.  We tested this design by mapping a pair of reference dipoles for which high-fidelity beam measurements and electromagnetic models are available.   From this first trial we draw several conclusions, discussed below.

\subsection{Mapping Efficiency}

The beam mapping experiment by Chang et al (2015) used a rectilinear grid and found difficulty efficiently sampling the beam with a uniform coverage in a reasonable amount of time.   In that case, the AUT was a dish with a fairly narrow field of view coverable with a level flight path. A level rectinilear grid becomes even more problematic when mapping a wide-field response.  Measuring the response to the horizon would require a very large flight pattern. Given the requirement of sampling the beam over the full half sphere we chose as our flight path the ring pattern HEALpix pixelization. This is a signficant advance over regular level grids as it makes possible sampling near to the horizon level without having to fly far away. However, it does little to ease the time required to acquire high angular resolution maps, which require dense flight patterns regardless of gridding scheme. Nor does it necessarily reduce the sample variance, which would require slower flight speeds to obtain more samples per gridded pixel. Both issues strongly motivate the need for longer flight times.  With a flight lifetime of 15 minutes for ECHO, each map reported here required four flights, which when performed with no technical issues, took about 100 minutes of total operational time since the downtime between flights to swap batteries adds an overhead of nearly 70\%.  A drone with a 30 minute flight time is possible with currently available technology and would make possible higher resolution maps in the same amount of operational time.

\subsection{Sources of Error}

We found that under most circumstances, the repeatability of multiple measurements in a gridded pixel had a typical standard deviation of 2\% after flagging on the telemetry data. However we see that the standard deviation increases as transmitter polarization becomes aligned with the receiver near the nulls of the beam patterns. In the null of the AUT beam, the standard deviation was about $\sim$6\%. One possible source of this instability is if the transmitter antenna were rotating in its harness. Turning to our model of the transmitter beam, we found a yaw RMS of  1.3\arcdeg{} is sufficient to cause a 6\% error. This error is likely to be a combination of actual yaw instability in the drone and instability of the monofilament mounting method.

\subsection{Areas for Improvement}

The sling mounting was chosen under the assumption that distance from the drone would improve the accuracy of the transmitter beam model, however subsequent modeling in CST has shown that changes in mounting distance yield small changes to the beam.  Given the instability of the sling method, investigation is currently under way into options for solid mounts closer to the drone body.  Furthermore, since measurement variance is consistent with excessive yaw in the system, we must ensure more stable flight and antenna mounting. Care must be taken to shield the drone magnetometer from the high currents feeding the prop motors and to properly tune the flight control loop settings to yield slow and stable turning.   Finally, due to increased inertia, stability increases with the size of the drone, hence larger drones may be better suited for the application than the one used in this work.

In the end, drone stability can only be refined so far. In most drone applications, unwanted motion is typically further suppressed with a self-stabilizing gimbal which is commonly used for photography and cinematography. Use of such a thing was avoided in this first iteration to minimize weight and to avoid the addition of more conductive components. As this instability is the largest source of error, the benefits from addition of a gimbal mechanism will likely outweigh the costs. 

Comparing our data with the electromagnetic model of the receiver dipole, we found that our data agreed with the model on one of the polarizations, but not the other.  This was true for both identically constructed test stations. As these stations serve as the underlying basis for the accuracy of both drone and Orbcomm systems, future work should focus on identifying the sources of these differences and improving the accuracy of the models.

Other areas for further development include extension across the 21~cm bands. Coverage of the reionization (100-200\,MHz) and BAO (400-800\,MHz) bands is possible with the current configuration. A larger antenna will be required for the cosmic dawn band 50-100\,MHz.  ECHO measurements are suitable to characterize the polarization characteristics of an AUT and we plan to explore cross-polarization measurements in future work.   The ECHO system  suffers from sensitivity issues at the poles due to our restriction that the transmitting dipole antenna is maintained parallel with the ground. Transmitter feeds that transmit both polarizations simultaneously and are gimbaled to point at the AUT are possible, particularly in the sub-meter bands, however more investigation is required to understand the limitations and tradeoffs.

\subsection{Overall Performance}

Is the quality of the mapping at the level required for precision foreground subtraction?   If we take as a worst-case, the foreground precision requirement to be 10,000:1, then our target precision is 0.01\%, while our current best demonstrated absolute accuracy is $\sim$1\%.  However, we note that the 0.01\% target is not well defined, since it is not well-studied how to propagate specifications from the required flux subtraction precision to the accuracy of the beam. \citet{Shaw2015_chimemmodes} have set a requirement to know the CHIME to beamwidth to 0.1\%, which for a gaussian beam translates to $\sim$1~dB error bars at a -40~dB level. Here we have demonstrated 2~dB at the -30~dB level.  More work is required to refine the target accuracy, as well as the best way to deliver and verify that accuracy, however it does appear to be achievable with modest performance improvements to the current system.

\section*{Open Project}
This project is an open-source platform which can be used and modified by anyone. The code used to plan flights and to reduce the data is available at \url{github.com/dannyjacobs/ECHO}; the data in this paper and the plotting scripts are available in this paper's repository \url{github.com/dannyjacobs/ECHO_paper1/}.

\acknowledgments
ECHO development is supported by a grant from the National Science Foundation AST program through award 1407646. D.C.J would like to acknowledge support by the NSF Astronomy and Astrophysics Fellowship Program through award 1401708.  We thank Rich Bradley and staff at the National Radio Astronomy Observatory, Green Bank and Andri Gretarsson and Embry Riddle Aeronautical Observatory for generously supporting this project with their time and equipment. This project makes use of the Astropy python library \citep{2013Robitaille_Astropy}.

\bibliographystyle{apj}

\end{document}